\documentclass[11pt]{article}
\usepackage{amssymb}
\usepackage{amsfonts}
\usepackage{latexsym}
\usepackage{graphicx}

\let\leq\leqslant
\let\geq\geqslant

\newcounter{proclineno}

\makeatletter
\def\@currentproclabel{}

\def\proclabel#1{\let\@currentproclabel\theproclineno\@bsphack\if@filesw%
{\let\thepage\relax\def\protect{\noexpand\noexpand\noexpand}%
\edef\@tempa{\write\@auxout{\string\newlabel{#1}{{\@currentproclabel}{\thepage}}}}%
\expandafter}\@tempa\if@nobreak \ifvmode\nobreak\fi\fi\fi\@esphack}%
\makeatother
\newtheorem{THEOREM}{Theorem}
\newenvironment{theorem}{\begin{THEOREM} \hspace{-.85em} {\bf .} }%
                        {\end{THEOREM}}
\newtheorem{LEMMA}{Lemma}
\newenvironment{lemma}{\begin{LEMMA} \hspace{-.85em} {\bf .} }%
                      {\end{LEMMA}}
\newtheorem{COROLLARY}{Corollary}
\newenvironment{corollary}{\begin{COROLLARY} \hspace{-.85em} {\bf .} }%
                          {\end{COROLLARY}}
\newtheorem{PROPOSITION}{Proposition}
\newenvironment{proposition}{\begin{PROPOSITION} \hspace{-.85em} {\bf .} }%
                            {\end{PROPOSITION}}
\newtheorem{PROPERTY}{\normalfont\textit{Property}}
\newenvironment{property}{\begin{PROPERTY} \hspace{-.85em} {\itshape .} \rm}%
                            {\end{PROPERTY}}
\newtheorem{DEFINITION}{Definition}
\newenvironment{definition}{\begin{DEFINITION} \hspace{-.85em} {\bf .} \rm}%
                            {\end{DEFINITION}}
\newtheorem{DEFINITIONL}[DEFINITION]{Definition}
\newenvironment{definitionl}{\begin{DEFINITIONL} \hspace{-.85em} {\bf .} \rm}%
                            {\end{DEFINITIONL}}
\newtheorem{EXAMPLE}{Example}
\newenvironment{example}{\begin{EXAMPLE} \hspace{-.85em} {\bf .} \rm}%
                            {\end{EXAMPLE}}
\newtheorem{ALGORITHM}{Algorithm}
\newenvironment{algorithm}{\begin{ALGORITHM} \hspace{-.85em} {\bf .} \rm}%
                          {\end{ALGORITHM}}
\newtheorem{PROCEDURE}{Procedure}
\newenvironment{procedure}{\begin{PROCEDURE} %
\setcounter{proclineno}{0}\hspace{-.85em} {\bf .} \rm}%
                          {\end{PROCEDURE}}
\newtheorem{REMARK}{Remark}
\newenvironment{remark}{\begin{REMARK} \hspace{-.85em} {\bf .} \rm}%
                            {\end{REMARK}}
\newtheorem{CLAIM}{Claim}
\newenvironment{claim}{\begin{CLAIM} \hspace{-.85em} {\bf .} \rm}%
                            {\end{CLAIM}}
\newtheorem{CLAIMEMPH}{Claim}
\newenvironment{claimemph}{\begin{CLAIMEMPH} \hspace{-.85em} {\bf .} }%
                            {\end{CLAIMEMPH}}
\newtheorem{HYPOTHESIS}{Hypothesis}
\newenvironment{hypothesis}{\begin{HYPOTHESIS} \hspace{-.85em} {\bf .}
    \rm}%
                            {\end{HYPOTHESIS}}
\newtheorem{HYPOTHESISL}[HYPOTHESIS]{Hypothesis}
\newenvironment{hypothesisl}{\begin{HYPOTHESISL} \hspace{-.85em} {\bf .} \rm}%
                            {\end{HYPOTHESISL}}
\newtheorem{FACT}{Fact}
\newenvironment{fact}{\begin{FACT} \hspace{-.85em} {\bf .} \rm}%
                            {\end{FACT}}
\newtheorem{FACTL}[FACT]{Fact}
\newenvironment{factl}{\begin{FACTL} \hspace{-.85em} {\bf .} \rm}%
                            {\end{FACTL}}

\newcommand{\thm}{\begin{theorem}}
\newcommand{\lem}{\begin{lemma}}
\newcommand{\pro}{\begin{proposition}}
\newcommand{\prop}{\begin{property}}
\newcommand{\dfn}{\begin{definition}}
\newcommand{\dfnl}{\begin{definitionl}}
\newcommand{\rem}{\begin{remark}}
\newcommand{\clm}{\begin{claim}}
\newcommand{\clme}{\begin{claimemph}}
\newcommand{\hypt}{\begin{hypothesis}}
\newcommand{\hyptl}{\begin{hypothesisl}}
\newcommand{\fct}{\begin{fact}}
\newcommand{\fctl}{\begin{factl}}
\newcommand{\xam}{\begin{example}}
\newcommand{\alg}{\begin{algorithm}}
\newcommand{\proc}{\begin{procedure}}
\newcommand{\cor}{\begin{corollary}}
\newcommand{\prf}{\noindent{\bf Proof.} }

\newcommand{\ethm}{\end{theorem}}
\newcommand{\elem}{\end{lemma}}
\newcommand{\epro}{\end{proposition}}
\newcommand{\eprop}{\bbox\end{property}}
\newcommand{\edfn}{\bbox\end{definition}}
\newcommand{\edfnl}{\end{definitionl}}
\newcommand{\erem}{\bbox\end{remark}}
\newcommand{\eclm}{\bbox\end{claim}}
\newcommand{\eclme}{\end{claimemph}}
\newcommand{\ehypt}{\bbox\end{hypothesis}}
\newcommand{\ehyptl}{\end{hypothesisl}}
\newcommand{\efct}{\bbox\end{fact}}
\newcommand{\efctl}{\end{factl}}
\newcommand{\exam}{\bbox\end{example}}
\newcommand{\ealg}{\end{algorithm}}
\newcommand{\eproc}{\end{procedure}}
\newcommand{\ecor}{\end{corollary}}
\newcommand{\eprf}{\bbox}

\newcommand{\beqn}{\begin{equation}}
\newcommand{\eeqn}{\end{equation}}

\newcommand{\bbox}{\vrule height7pt width4pt depth1pt}

\def \opt{\mathop{\mathrm{opt}}}

\def\ids{\textsc{min independent dominating set}}

\title{\textbf{Fast algorithms for \ids{}}}

\author{N. Bourgeois \hspace*{1cm} B. Escoffier \hspace*{1cm} V.~Th.~Paschos \vspace*{2mm} \\ 
LAMSADE, CNRS FRE~3234 and Universit\'e Paris-Dauphine, France \\
\texttt{\{bourgeois,escoffier,paschos\}@lamsade.dauphine.fr}}

\begin{document}

\maketitle

\begin{abstract}

We first devise a branching algorithm that computes a minimum independent dominating set on any graph with running time~$O^*(2^{0.424n})$ and polynomial space. This improves the~$O^*(2^{0.441|V|})$ result by (S.~Gaspers and M.~Liedloff, \textit{A branch-and-reduce algorithm for finding a minimum independent dominating set in graphs}, Proc.~WG'06). We then show that, for every $r \geqslant 3$, it is possible to compute an $r - (({r-1})/{r})\log_2r$-approximate solution for \ids{} within time~$O^*(2^{n\log_2r/r})$.

\end{abstract}

\section{Introduction}\label{introduction}

An independent set in a graph~$G(V,E)$ is a vertex subset $S
\subseteq V$ such that for any $(v_i,v_j) \in S \times S$,
$(v_i,v_j) \notin E$. An independent dominating set is an
independent set that is maximal for inclusion. \ids{} is known to be
\textbf{NP}-hard~\cite{gj}. Consequently, it is extremely unlikely
that a polynomial algorithm could ever be designed solving it to
optimality. Unfortunately, this problem is also very badly
approximable since no polynomial algorithm can approximately solve
it within ratio~$|V|^{1-\epsilon}$, for any $\epsilon>0$, unless
$\textbf{P} = \textbf{NP}$~\cite{halmmis}.

Except polynomial approximation, another way to cope with
intractability of \ids{} (as well as of any other \textbf{NP}-hard
problem) is by designing algorithms able to solve it to optimality
with worst-case exponential running time as low as possible. Since
\ids{} can be trivially solved in~$O(2^{|V|}p(|V|))$ by simply
enumerating all the subsets of~$V$, the stake of such a research
issue is to solve it within~$O(2^{c|V|}p(|V|))$, where~$c$ is a
constant lower than~1 and~$p$ is some polynomial. Since in
comparison with the slightest improvement of~$c$,~$p$ is non
relevant, we use from now on notation~$O^*(2^{c|V|})$ to measure the
complexity of an algorithm, this notation meaning that
multiplicative polynomial factors are ignored.

For \ids{}, trivial~$O^*(2^{|V|})$ bound has been initially broken
by~\cite{joyapa88} down to~$O^*(3^{|V|/3})$ using a result
by~\cite{moomo}, namely that the number of maximal (for inclusion)
independent sets in a graph is at most $3^{|V|/3}$. Then, obviously,
it is possible to compute a minimum independent dominating set
in~$O^*(3^{|V|/3})=O^*(2^{0.529|V|})$ using polynomial space.
This result has been dominated by~\cite{galie06} where using a
branch~\& reduce technique an algorithm optimally solving \ids{}
with running time~$O^*(2^{0.441|V|})$ is proposed.

In this paper, we first devise a branching algorithm that can find a minimum independent dominating set on any graph with running time~$O^*(2^{0.424|V|})$ and polynomial space. We then show that, for every $r \geqslant 3$, it is possible to compute an $r$-approximate solution for \ids{} within time~$O^*(2^{|V|\log_2r/r})$. Finally, we improve this ratio down to $r - (({r-1})/{r})\log_2r$ and prove that it can be achieved always in time~$O^*(2^{|V|\log_2r/r})$.

In what follows, given a graph~$G(V,E)$ and a vertex $v \in V$, the neighborhood~$N(v)$ of~$v$ is the set of vertices that are adjacent to~$v$, and $N[v] = N(v) \cup \{v\}$ will be called the closed neighborhood of~$v$. For the degree of~$v$, we use the notation $d(v)=|N(v)|$. For any subset $H \subset V$, we denote by~$G[H]$ the subgraph of~$G$ induced by~$H$. For $v \in H$, for some subset $H$, we denote by~$d'_H(v)$ the degree of~$v$ in~$G[H]$ or, if it is clear by the context, we denote it by~$d'(v)$. For convenience, we set $N[H] = \{N[v]: v\in H\}$. We use~$\delta$ and~$\Delta$ to denote the minimum and maximum degree of~$G$, respectively. For simplicity, we set $n=|V|$ and $m=|E|$;~$T(n)$ stands for the maximum running time an algorithm requires to solve \ids{} in a graph containing at most~$n$ vertices.


We conclude this introduction by a remark that has to be taken into
account when operating sequentially several branchings, in order to
get a sound bound on the complexity. Let us call a branching
``single'' if given a vertex~$v$, one has to decide what vertex
in~$N[v]$ dominates~$v$. A ``multiple'' branching is a decision
tree, nodes of which correspond to simple branchings.

Branch \&~reduce-based algorithms have been used for decades, and a
classical analysis of their running times leading to worst case
complexity upper bounds is now well-known. Let~$T(n)$ be an upper
bound on the running time of the algorithm for an instance of size
at most~$n$. If we now that computing a solution on an instance of
size~$n$ amounts to computation of a solution on a sequence of~$p$
instances of respective sizes $n-k_1, \ldots, n-k_p$, we can write:
\begin{equation}\label{model}
T(n) \leq \sum_{i\leq p}T\left(n-k_i\right)+q(n)
\end{equation}
for some polynomial $q$. The running time~$T(n)$ is bounded
by~$O^*(c^n)$, where~$c$ is the largest positive real root of: $1 =
\sum_{i\leq p}x^{-k_i}$. This root is often called the contribution
of the branching to overall complexity factor, or the complexity
factor of the branching. Note that in the remainder of the article,
for simplicity, we will omit to precise the additive polynomial term
$q(n)$ in recurrence relations as \ref{model}

 Now, it is possible that there is not only a single
recurrence as in~(\ref{model}), but several ones, depending on the
instance. For example, either $T(n) \leq 2T(n-2)$ or $T(n) \leq
7T(n-7)$. Fortunately, as far as single branchings are concerned,
the analysis is very simple: the running time is never greater than
what is needed to solve an instance where at every step we make a
branching that has the highest possible complexity factor, i.e., the
greatest solution of~(\ref{model}).

On the other hand, multi-branching analysis may encounter some
problems. Indeed, this rule is not true any more when we make
multiple branchings. Multiple branching is based upon a very simple
idea: instead of choosing between some branchings, we choose between
some sequences of branching, such as: ``if one adds~$a$ to the
solution, then one knows that~$b$ has degree~$3$ in the remaining
graph and one can make a very good branching on it~\dots''. The
efficiency of such a sequence can be measured with inequalities very
close to~(\ref{model}). If for instance, the first branching allows
us to consider two graphs with, say,~$2$ fewer vertices, and we know
that one of these graphs is good enough to allow us to consider
three graphs with~$7$ fewer vertices, we can state the following
recurrence:
$T(n) \leq T(n-2)+3T(n-2-7)$.

\begin{figure}[htb]
\begin{center}
\includegraphics{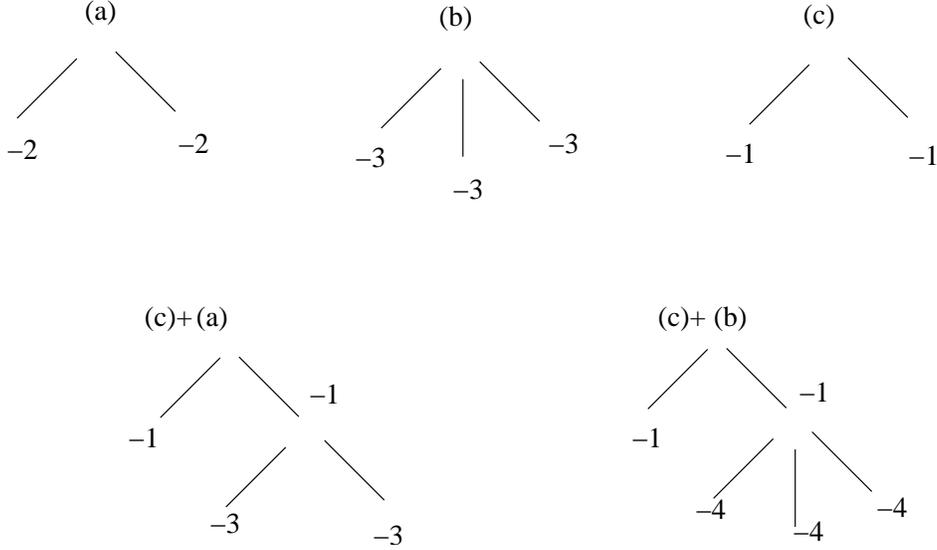}
\caption{A multiple branching that can be misleading when dealing with the evaluation of the highest complexity factor.}\label{multibr}
\end{center}
\end{figure}

The problem arises when we have several choices for one of these
``elementary'' branchings. Unfortunately, we cannot simply try the
one with the higher complexity factor, because it is possible that a
locally smaller factor leads to a globally higher one, see for
example Figure~\ref{multibr}. Although branching~(a) has a higher
complexity factor than branching~(b)~($1.443$ instead of~$1.415$),
the branching (c)+(a) is better than (c)+(b)~($1.659$ and~$1.696$,
respectively).

In what follows, if a branching is never used in a combination with another one, we simply give the inequality with the highest factor. Otherwise, we give all the relevant inequalities.

\section{General recurrence}\label{generalreccurence}

Following an idea by~\cite{galie06}, we partition the graph into
``marked'' and ``free'' vertices. Marked vertices are those that
have already been disqualified from belonging to optimum, but are
not dominated yet. In other terms, we generalize the problem at hand
in the following way: given a subset $W \subseteq V$ ($W$ is the set
of free vertices), find the minimum independent set in~$W$ that
dominates~$V$. Notice that, without further hypothesis on~$W$, this
problem may have no solution (for instance, when $V \setminus W$
contains a vertex and its whole neighborhood); in this case we
consider $\opt(G,W)=\infty$.

In what follows we say that two vertices~$v$ and~$u$ are equivalent if $N[u] = N[v]$. In this case, we can remove from the graph one of them, a marked one if any, otherwise at random.
\begin{lemma}\label{geniq}
Let~$\delta$ be the minimum degree of~$G(V,E)$, and~$r$ be the
maximum number of marked vertices in~$N[v]$, for any~$v$ such
that~$d(v)=\delta$. Then:
$$
T(n) \leq (\delta+1-r)T(n-\delta-1)
$$
Furthermore, when branching on some neighbor of~$v$, we can mark all other neighbors that have already been examined.
\end{lemma}
\prf We can always suppose that $\delta \geq 1$, since (not marked)
isolated vertices must be added to the solution. Let~$v$ be a vertex
such that $d(v)=\delta$. Our solution has to dominate~$v$, so at
least one vertex of~$N[v]$ must belong to, and this vertex must be a
free one. Furthermore, if some vertex~$u$ belongs to optimum, its
neighbors do not so and, consequently, they are dominated. Hence, we
get the following recurrence:
\begin{equation}
\opt(G) = 1 + \min_{u \in W\cap N[v]}\left\{\opt(G[V\setminus N[u]])\right\} \label{genrec}
\end{equation}
By hypothesis, $d(u) \geq \delta$, so we get the inequality claimed.

\begin{figure}[htb]
\begin{center}
\includegraphics{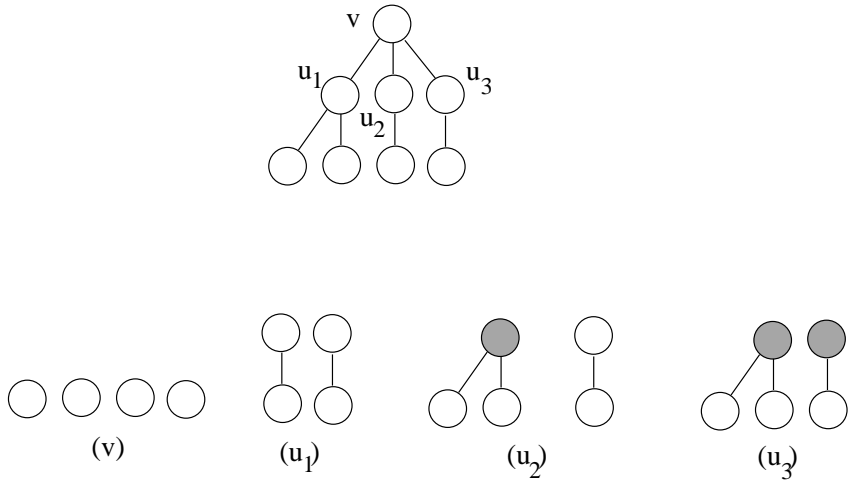}
\caption{The four branches
$v;\bar{v}u_1;\bar{v}\bar{u_1}u_2;\bar{v}\bar{u_1}\bar{u_2}u_3$.}\label{finlemma1}
\end{center}
\end{figure}

Remark that the order we use to examine neighbors~$u_i$, $i = 1,
\ldots, \delta$ of~$v$ is important, since the sequence of choices
is not ``$v ; u_1 ; u_2 ; \ldots ;u_{\delta}$'' but ``$v ;\bar{v}u_1
; \bar{v}\bar{u_1}u_2 ; \ldots ;\bar{v}\bar{u_1} \ldots
\bar{u_{\delta-1}}u_{\delta}$'' where for a vertex $u$,~$\bar{u}$
means ``not~$u$'' and we assume that~$u_i$'s are ordered in
increasing degree order. Figure~\ref{finlemma1} illustrates the
proof of the lemma.~\eprf

Note that complexity of branching is decreasing with~$\delta$, for $\delta \geq 2$.
So, a straightforward idea is to perform a fine analysis on graphs of low minimum degree. Formally, the algorithm we propose works as follows:
\begin{itemize}
\item if there exists a marked vertex of degree~$3$ or less, or a vertex which is adjacent to
only one free vertex, make a branching according to what is
described in Section~\ref{marked};
\item otherwise, pick a vertex of minimum degree, and branch as described in Section~\ref{lowdegree}.
\end{itemize}
In our analysis of the running time, we adopt a measure and conquer
approach. More precisely, we do not count in the measure the marked
vertices of degree at most 2 (they receive weight 0), and we count
with a weight $w=0.2$ the marked vertices of degree 3. The other
vertices receive weight 1. We get recurrences on the time $T(p)$
required to solve instances of weight $p$, where the weight of an
instance is total weight of the vertices in the graph. Since
initially $p=n$ we get the running time as a function of $n$. This
is valid since when the total weight is $p=0$, there are only marked
vertices in the graph and we can solve the problem (there is no
solution). Note that this way of measuring progress is introduced in
order to simplify the branching analysis (we could obtain a similar
result without measure and conquer but with a deeper analysis).

\section{Branching on marked vertices or vertices which are adjacent to only one free vertex}\label{marked}

As a preliminary remark note that if there is an edge between two
adjacent marked vertices, we can remove this edge. This removal does
not increase our complexity measure $p$.

We branch as follows: either there is a marked vertex of degree at most~2 (Lemma~\ref{red2}), or a (free) vertex which is adjacent to only one free vertex (Lemma~\ref{deg1}) or a marked vertex of degree~3 (Lemma~\ref{red3}). If there are no such vertices, then go to Section~\ref{lowdegree}.
\begin{lemma}\label{red2}
Assume some vertex of degree at most~$2$ is marked. Then either the
algorithm ends (there is no solution), or we can remove at least one
vertex without branching, or $T(p) \leq T(p-2)+T(p-4)$; this
branching contributes to the overall complexity with a factor
$\lambda \leq 1.2721=2^{0.348}$.
\end{lemma}
\prf If there is a marked vertex $v$ of degree 0, then  then
$\opt(G)=\infty$.

If $v$ has degree~$1$ we add its neighbor $u$ to the solution and we
reduce the current instance's weight by~1 without branching.

Suppose now that~$v$ is marked and adjacent to~$u_1,u_2$ (which are free). Then:
$$
\opt(G) = 1 + \min\left\{\opt\left(G\setminus N\left[u_1\right]\right),opt\left(G\setminus N\left[u_2\right]\right)\right\}
$$
If both~$u_1$ and~$u_2$ are adjacent to at least 2 free vertices,
then $T(p) \leq 2T(p-3)$, that is better than the result claimed. If
some~$u_i$ is adjacent only to marked vertices, we must add it to
the optimum, decreasing~$p$ by~$1$ without branching. Otherwise,
$u_1$ is adjacent to only one free vertex~$t_1$ and~$u_2$ is adjacent to at least one free vertex~$t_2$. One of the following
situations occurs:
\begin{itemize}
\item If~$u_1$ and~$u_2$ are adjacent: if $u_2$ is adjacent to two other free vertices $t_2$ and $t_3$, then
if we take $u_1$ we reduce $p$ by 2, if we take $u_2$ we reduce it
by 4. If $u_2$ is adjacent to only one vertex $t_2$, then taking
$u_1$ is interesting only if we take $t_1$. Hence, either we take
$u_1$ (and $t_1$) and $p$ reduces by 3, or we take $u_2$ and $p$
reduces by 3.
\item Otherwise, if~$t_1=t_2$, the only possibility to have both~$u_1$ and~$v$
dominated is to add~$u_1$ to the solution; thus we reduce the
current instance without branching.
\item Finally, if $N[u_1]\cap N[u_2]=\{v\}$, we branch on~$v$; when we add~$u_2$ to the solution, we must add~$t_1$  too,
in order to dominate~$u_1$; this leads to $T(p) \leq T(p-2)+T(p-4)$.~\eprf
\end{itemize}
Now, we suppose that the graph does not contain any marked vertex of degree at most~2.
\begin{lemma}\label{deg1}
If there exists $v \in V$ such that $N(v)\cap W=\{u\}$, then
$T(p)\leq 2T(p-(2+2w))$, and the complexity factor induced is
$\lambda \leqslant 1.3349=2^{0.417}$.
\end{lemma}
\prf Let~$u$ be the only free neighbor of~$v$. If $d(u)=1$, we can
add $v$ to the solution and discard~$u$ without branching.
Otherwise, if $u$ or $v$ is adjacent to at least two vertices of
weight 1,~removing $N[u]$ (or $N[v]$) reduces $p$ by at least~$3$
vertices. Taking either $v$ or $u$ gives $T(p)\leq T(p-2)+T(p-3)$.
The only remaining situation occurs when all the other neighbors of
both $u$ and $v$ are marked and of degree 3. If there is only one
such vertex, then $u$ and $v$ are equivalent and we don't need to
branch. Otherwise, when taking $u$ or $v$ these marked vertices of
degree 3 either are removed or become of degree 2, hence $T(p)\leq
2T(p-(2+2w))$.~\eprf
\begin{lemma}\label{red3}
Assume some vertex $v$ of degree~$3$ is marked. Then in the worst
case $T(p)\leq 2T(p-(3+w))+T(p-(5+w))$, and the complexity factor
induced is $\lambda \leqslant 1.3409=2^{0.424}$.
\end{lemma}
\prf Let $\{u_1,u_2,u_3\}$ the three neighbors of $v$.  One of the
following situations occur:
\begin{enumerate}
\item If each~$u_i$ is adjacent to at least~3 free vertices, then, by taking either~$u_1$, $u_2$ or~$u_3$ we get three branches of weight at most $p-(4+w)$.
\item If say~$u_1$ is adjacent to two free vertices, then we branch on~$u_1$: if we take~$u_1$ we reduce~$p$ by at least $3+w$. Otherwise, we do not take~$u_1$. In this case~$u_1$ is marked and we can remove the edges between~$u_1$ and the marked vertices. Hence,~$u_1$ and~$v$ are marked and have degree at most~2. Then,~$p$ reduces by $1+w$. But a further branching on a marked vertex of degree at most~2 (see Lemma~\ref{red2}) allows either to reduce~$p$ by~1, or creates two branches of weight at most $1+w+2=3+w$ or $1+w+4=5+w$. In the worst case, $T(p)\leq 2T(p-(3+w)) + T(p-(5+w))$.~\eprf
\end{enumerate}

\section{Branching on vertices of minimum degree}\label{lowdegree}

Now, we suppose that the graph does not contain any marked vertex of
degree at most~3, and that every vertex is adjacent to at least two
free vertices. Then, we branch on the vertex of minimum degree. If
this minimum degree is at least~6, then the branching given in
Section~\ref{generalreccurence} gives a sufficiently low running
time. We distinguish in the following lemmas the different possible
values of the minimum degree. Let us start with two preliminary
remarks. \rem\label{remneighbor} When branching on a vertex of
minimum degree~$\delta$, we can always assume that it is adjacent to
at least one vertex of degree at least $\delta+1$. Notice that the
degree of a vertex never increases. Then, the situation where the
graph is $\delta$-regular arrives at most once (even in case of
disconnection). Thus, we make only a finite number of ``bad''
branchings, fact that may increase the global running time only by a
constant factor. In particular, if $\delta=5$, the branching given
in Section~\ref{generalreccurence} gives $T(p)\leq 5T(p-6)+T(p-7)$
leading to a complexity factor $1.3384=2^{0.421}$.~\erem
\rem\label{rembranch} Suppose that we branch on a vertex $v$ which
has a neighbor $u_1$ adjacent to at most 3 free neighbors. Let us
consider the branch where we take $u_k$ not adjacent to $u_1$ (for
$2\leq k \leq d(v)$). In this branch $u_1$ is marked.
\begin{itemize}
\item If $N_W(u_1)\subseteq N_W(u_k)$, then we cannot take~$u_k$ and this branch is useless.
\item
If there is only one vertex~$t$ in $N_W(u_1)\setminus N_W(u_k)$: in this branch we have to take~$t$ and then we remove at least $d(u_k)+3$ vertices ($d(u_k)+1$ by taking~$u_k$, and~$t$ and~$u_1$ by taking~$t$).
\item
Otherwise~$u_1$ has two other free neighbors~$t_1, t_2$ which are not in~$N(u_k)$: in this branch we create a marked vertex ($u_1$) of
degree 2 and hence reduce $p$ by $d(u_k)+2$. Thanks to
Lemma~\ref{red2}, a further branching on the marked vertex of degree
at most 2 created gives two
branches where $p$ reduces by at least $d(u_k)+2+2$ and $d(u_k)+2+4$. 
\end{itemize}
In all, either we have one branch with a reduction of  $d(u_k)+3$,
or two branches with $d(u_k)+4$ and  $d(u_k)+6$ (the latter will
always be the worst case).~\erem
We are ready now to state the following theorem, that is the main result of the paper.
\begin{theorem}\label{followingtheorem}
\ids{} can be solved in~$O^*(2^{0.424n})=O^*(1.3413^n)$, using
polynomial space.
\end{theorem}
The proof of Theorem~\ref{followingtheorem} is immediate consequence of putting together Lemmata~\ref{deg2}, \ref{deg3} and~\ref{deg4} below that deal with the cases of minimum degree~2, 3 and~4, respectively.
\begin{lemma}\label{deg2}
If there exists $v \in V$ such that $N(v)=\{u_1,u_2\}$, then in the
worst case we get $T(p)\leq T(p-3)+T(p-4)+T(p-6)+T(p-8)$ and the
complexity factor induced is $\lambda \leqslant 1.3384=2^{0.421}$.
\end{lemma}
\prf One of the following situations occur (note that either $u_1$
or $u_2$ has degree at least 3):
\begin{enumerate}
\item If $d(u_2)\geq 4$ and $d(u_1)\geq 3$ then by taking either~$v$, or~$u_1$, or~$u_2$, we get three branches of weight at most $p-3$, $p-4$ and
 $p-5$.
\item If $d(u_2)\geq 4$ and $d(u_1)=2$ then, when we take~$u_2$~($u_1$ having already been discarded), we must also add the only remaining
neighbor~$t$ of~$u_1$ to the solution. Thus, thanks to
Remark~\ref{rembranch} (first and second item), we get $T(p) \leq
2T(p-3)+T(p-7)$.
\item If~$u_1$ and~$u_2$ have degree~3
and if there are adjacent: let $t$ the third neighbor of $u_1$ ($t$
is not adjacent to $u_2$ otherwise $u_1$ and $u_2$ are equivalent
and we can remove one of them). When taking $v$, we can take also
$t$ since if we don't take $t$ it is useless to take $v$. Hence, we
get three branches, each of weight at most $p-4$ (even better on the
first branch actually).
\item If $d(u_1)=3$, then $u_1$ and $u_2$ are not adjacent
(either because the case has been dealt before, or because
$d(u_2)=2$ and $u_2$ would be equivalent to $v$), then by branching
on $v$, either we take $v$ (weight at most $p-3$) or we take $u_1$
(weight at most $p-4$) or we take $u_2$ and we don't take $v$ and
$u_1$. According to Remark~\ref{rembranch}, this last choice reduces
$p$ either by $d(u_2)+3=5$, or gives birth to two branches of weight
at most $p-6$ and $p-8$.~\eprf
\end{enumerate}
\begin{lemma}\label{deg3}
If there exists $v \in V$ such that $N(v)=\{u_1,u_2,u_3\}$, then in
the worst case we get $T(p)\leq T(p-4)+3T(p-5)+T(p-6)+T(p-8)$ and
the complexity factor induced is $\lambda \leqslant
1.3413=2^{0.424}$.
\end{lemma}
\prf
If there exists such a vertex $v$ which is marked, then we only have
to consider 3 branches where $p$ reduces by at least 4: $T(p)\leq
3T(p-4)$. This is the same if one of the neighbors of $v$ is marked.

Now we consider that neither $v$ nor the $u_i$'s are marked. If
$4\leq d(u_i)$, $i=1,2,3$, by branching on $v$ we get one branch of
weight $p-4$ and 3 branches of weight at most $p-5$. Now, we
consider that $u_1$ has degree 3 and $u_3$ has degree at least 4.
Note that $u_1$ cannot be adjacent to $u_2$ and $u_3$ otherwise it
is equivalent to $v$. We consider the three following  cases: either
there are two edges in $N(v)$, or 1 or zero.
\begin{enumerate}
\item There are two edges in $N(v)$. Then wlog., $u_3$ is adjacent to both $u_1$ and $u_2$
(and $u_1$ is not adjacent to $u_2$). We get four branches of weight
at most $p-4$, $p-4$, $p-(d(u_2)+2)$ (since $u_1$ becomes marked and
of degree at most 2) and $p-(d(u_3)+1)$. In the third branch, thanks
to Remark~\ref{rembranch}, either we remove one more vertex or we
get two branches of weight at most $p-(d(u_2)+4)\leq p-7$ and
$p-(d(u_2)+6)\leq p-9$.
\begin{itemize}
\item If $d(u_3)\geq 5$: in the worst case $T(p)\leq 2T(p-4)+T(p-7)+T(p-9)+T(p-6)$.
\item Otherwise $d(u_3)=4$. 
Let $t$ be the fourth neighbor of $u_3$. In the branch we take $v$
we can take $t$ (indeed, if we don't take $t$ it is useless to take
$v$, taking $u_3$ is always better). Hence, we get one more vertex
deleted when taking $v$ (even more actually) and get the following
recurrence: $T(p)\leq T(p-5)+T(p-4)+2T(p-5)$.
\end{itemize}
\item Otherwise, there is at most one edge in $N(v)$.
\begin{enumerate}
\item If there is a triangle of vertices of degree 3 $v,u_1,u_2$, let
$t_1$ and $t_2$ be the third neighbors of $u_1$ and $u_2$.  If~two
vertices among $u_3$, $t_1$ and~$t_2$ are equal or adjacent, either
two vertices in the triangle are equivalent (case equal) or one
vertex in the triangle $v,u_1,u_2$ must belong to the solution,
otherwise one of them would not be dominated (case adjacent); hence,
$T(p) \leq 3T(p-4)$. Finally, if~$t_1$,~$t_2$ and $u_3$ are distinct and non adjacent,
set $\Gamma'=N(t_1) \cup N(t_2) \cup N(t_3)\setminus
\{v,t_1,t_2,u_1,u_2,u_3\}$; either we take one vertex of the
triangle, or we have to take~$t_1$,~$t_2$ and $u_3$:
$T(p) \leq 3T(p-4)+T(p-6-|\Gamma'|)$.
If $|\Gamma'|\geq 4$, $T(p) \leq 3T(p-4)+T(p-10)$ leading to
$\lambda \leq 2^{0.417}$. If $|\Gamma'|= 3$, in this case $u_3$ has
degree 4 and $\Gamma'\subset N(u_3)$. Then, we branch as follows:
either we take $u_3$ and remove 9 vertices, or we take $v$ and
remove 4 vertices, or we mark $u_3$ and $v$. By removing the edge
between them, they have respectively degree 3 and 2. Hence $T(p)\leq
T(p-9)+T(p-4)+T(p-(2-w))$.
\item Otherwise, if the only edge in $N(v)$ is $(u_2,u_3)$ we get $T(p)\leq T(p-4)+T(p-4)+T(p-5)+T(p-6)$, but
in the two last branches, thanks to Remark~\ref{rembranch}, either
we remove one more vertex or we get two branches of weight reduced
by 2 and 4. In the worst case, we get: $T(p)\leq
2T(p-4)+T(p-7)+T(p-9)+T(p-8)+T(p-10)$.
\item If there is an edge $u_1,u_2$ with $d(u_2)\geq 4$ (otherwise this is the triangle case) we get $T(p)\leq
T(p-4)+T(p-4)+T(p-5)+T(p-6)$, but in the last branch, again thanks
to Remark~\ref{rembranch}, we get in the worst case $T(p)\leq
2T(p-4)+T(p-5)+T(p-8)+T(p-10)$.
\end{enumerate}
\item  Finally, if there is no edge in $N(v)$: if two $u_2$ and $u_3$ have degree at least 4, then we
get $T(p)\leq T(p-4)+T(p-4)+T(p-6)+T(p-6)$ (indeed in the last two
branches $u_1$ is marked and of degree at most 2). If say $u_2$ has
degree 3, then we get $T(p)\leq T(p-4)+T(p-4)+T(p-5)+T(p-6)$, but in
the two last branches, thanks to Remark~\ref{rembranch}, either we
remove one more vertex or we get two branches of weight reduced by 2
and 4. In the worst case, we get: $T(p)\leq
2T(p-4)+T(p-7)+T(p-9)+T(p-8)+T(p-10)$.~\eprf
\end{enumerate}
\begin{lemma}\label{deg4}
If there exists $v \in V$ such that $N(v)=\{u_1,u_2,u_3,u_4\}$, then
in the worst case we get $T(p) \leq 4T(p-5)+T(p-9)$
with a contribution to the overall complexity factor bounded above
by~$1.3394=2^{0.422}$.
\end{lemma}
\prf We consider that~$u_4$ has degree at least~5. If at least~3~$u_i$'s have degree at least~5, then $T(p) \leq 2T(p-5)+3T(p-6)$. Now, we consider that~$u_1$ and~$u_2$ have degree~4.

Suppose first that the~$u_i$'s of degree~4 are not adjacent. In the branch we take~$u_2$,~$p$ reduces by $6-w$ (since~$u_1$ is marked and of degree at most~3). Then either~$u_3$ has degree at least~5 and then $T(p)\leq 2T(p-5)+T(p-(6-w))+2T(p-6)$, or~$u_3$ has degree~4 and in this case, in the branch we take~$u_3$,~$p$ reduces by $5+2(1-w)$: $T(p)\leq 2T(p-5)+T(p-(6-w))+T(p-(7-2w))+T(p-6)$.


If $u_1$ and $u_2$ are adjacent. When $v,u_1$ and $u_2$ are marked,
they become of degree 2. Then:
\begin{itemize}
\item Either $u_1$
and $u_2$ are not adjacent to $u_3$ and we get $T(p)\leq
3T(p-5)+T(p-7)+T(p-6)$.
\item Or $u_1$ is not adjacent to $u_3$ and
$u_2$ is not adjacent to $u_4$ and we get $T(p)\leq
3T(p-5)+T(p-6)+T(p-7)$.
\item Otherwise both~$u_1$ and~$u_2$ are adjacent to~$u_3$ and not to~$u_4$. If~$u_3$ has degree at least~5, $T(p)\leq 3T(p-5)+T(p-6)+T(p-7)$. Otherwise,~$u_4$ is not adjacent to any of the~$u_i$'s (if not, one would be equivalent to~$v$). Then, we get~4 branches of weight $p-5$ and in the last branch the~$u_i$'s are marked and are of degree at most~2, hence~$p$ reduces by~$9$: $T(p)\leq 4T(p-5)+T(p-9)$.~\eprf
\end{itemize}

\section{Approximation of \ids{} by moderately exponential algorithms}

As we have mentioned in Section~\ref{introduction}, for any
$\varepsilon>0$, \ids{} is inapproximable within ratio
$O(n^{1-\varepsilon})$ unless $\textbf{P} = \textbf{NP}$. On the
other hand, it is easy to see that any maximal independent set
guarantees a ratio at most $\Delta +1$. In this section, we devise
algorithms achieving ratios much better than~$O(n)$, i.e.,
``forbidden''  in polynomial time, and with running times that,
although exponential, are better than the running time of exact
computation for \ids{}. Our results are based upon the following
lemma by~\cite{byskov}.
\begin{lemma}\label{smallids}~(\cite{byskov})
For any $k \geq 3$, it is possible to compute any independent dominating set (i.e., maximal independent set) of size at most~$n/k$ with running time~$O^*(k^{n/k})$. This bound is tight.
\end{lemma}
\begin{proposition}\label{alg1}
For any $r \geq 3$, it is possible to compute an $r$-approximation of \ids{} with running time~$O^*(2^{n\log_2r/r})$.
\end{proposition}
\prf We run the branching algorithm leading to Lemma~\ref{smallids}.
If it finds some minimum independent dominating set, our algorithm
returns it; otherwise, $\opt(G) > n/r$, where~$\opt(G)$ denotes the
size of a minimum independent dominating set, and the algorithm
returns some arbitrary independent dominating set. In the first
case, the algorithm needs time $O^*(r^{n/r}) = O^*(2^{n\log_2r/r})$
and computes an optimal solution; in the second case, any maximal
independent set is an $r$-approximation and such a set is computed
in polynomial time.~\eprf

The following proposition further improves the result of Proposition~\ref{alg1}.
\begin{proposition}\label{alg2}
For any $r \geq 3$, it is possible to compute an approximation of \ids{} with running time $O^*(2^{n\log_2r/r})$ and approximation ratio $r- ({(r-1)}/{r})\log_2r$.
\end{proposition}
\prf
As previously, we first compute every independent dominating set of size~$n/r$ or less. If such sets exist, we return one of those with minimum size. Otherwise, we partition~$V$ into $l=r/\log_2r$ subsets $V_1, \ldots, V_l$, of size~$n\log_2r/r$, and we initialize~$S$ with some independent dominating set. Then, for $j \leq l$, we run the following procedure:
\begin{itemize}
\item for any $H \subset V_j$: if~$H$ is an independent set, compute an independent dominating set~$S_H$ in $G[V\setminus N[H]]$; if $|S| \geq |H \cup S_H|$, set $S = H \cup S_H$;
\item return~$S$.
\end{itemize}
Obviously,~$S$ is an independent dominating set. The algorithm has examined $l \times 2^{n/l}$ subsets, that concludes the running time claimed.

Fix some optimal solution~$S^*$ for \ids{}. Since $S^*$ is maximally
independent, $\cup_{i\leq l}N(V_i \cap S^*) = V \setminus S^*$ and
we get (I.S. stands for independent set):
\begin{eqnarray*}
\frac{|S|}{\opt(G)} &\leq & \min_{j \leq l}\min_{H \mbox{ {\footnotesize I.S. of} } V_j}\left\{\frac{n-|N(H)|}{\opt(G)}\right\} \;\; \leq \;\; \min_{j \leq l}\left\{\frac{n-\left|N\left(V_j \cap S^*\right)\right|}{\opt(G)}\right\} \;\; \leq \;\; \frac{n-\frac{n-\opt(G)}{l}}{\opt(G)} \\
&\leq& \frac{\log_2 r}{r} + r -\log_2r
\end{eqnarray*}
that completes the proof of the proposition.~\eprf

\begin{table}[htb]
\begin{center}
\begin{tabular}{c || c c c c c c c}
\hline\hline
{Ratio} & 2 & 3 & 4 & 5 & 10 & 20 & 50 \\
\hline\hline
Proposition~\ref{alg1} & & $1.4423^n$ & $1.4143^n$ & $1.3798^n$ & $1.2590^n$ & $1.1616^n$ & $1.0814^n$ \\
Proposition~\ref{alg2} & $1.4403^n$ & $1.3870^n$ & $1.3419^n$ & $1.3077^n$ & $1.2130^n$ & $1.1398^n$ & $1.0749^n$ \\
\hline\hline
\end{tabular}
\caption{Tradeoffs between running times and ratios derived by Propositions~\ref{alg1} and~\ref{alg2}.}\label{tradeoff}
\end{center}
\end{table}

Tradeoffs between running times and ratios for some values of the ratios are displayed, for both propositions in Table~\ref{tradeoff}. Recall that the exact algorithm given above runs in~$O^*(1.3416^n)$.

\bibliographystyle{plain}

\end{document}